\documentclass[useAMS,usenatbib]{mn2e}
\usepackage{times}
\usepackage{graphicx}
\title[A New Method of the Corotation Radius Evaluation]{
A New Method of the Corotation Radius Evaluation in our Galaxy}
\author[Acharova et. al.]
{ I. A. Acharova,$^{1}$\thanks{E-mail:
iaacharova@sfedu.ru (IAA); unmishurov@sfedu.ru (YuNM)} 
Yu. N. Mishurov,$^{1,2}$
M.R.Rasulova $^{1}$  
\\
$^{1}$ Department of Physics of Cosmos, Southern Federal University, 
5 Zorge, Rostov-on-Don, 344090, Russia \\
$^{2}$ Special Astrophysical Observatory of Russian Academy of Sciences, 
N.Arkhyz, Karachaevo-Cherkessia, Russia\\
}

\begin{document}

\date{Accepted 2011 xxxx. Received 2011 xxxx; in original form 2011 xxxx}

\maketitle

\pagerange{\pageref{firstpage}--\pageref{lastpage}} \pubyear{2011}

\label{firstpage}

\begin{abstract}
We propose a new method for determination of the rotation velocity of the galactic spiral density waves, correspondingly, the corotation radius, $r_C$, in our Galaxy by means of statistical analysis of radial oxygen distribution in the galactic disc derived over Cepheids. The corotation resonance happens to be located at $r_C \sim 7.0 - 7.6 $ kpc, depending on the rate of gas infall on to the galactic disc, the statistical error being $\sim 0.3 - 0.4$ kpc. Simultaneously, the constant for the rate of oxygen synthesis in the galactic disc was determined. 
       
We also argue in favour of a very short time-scale formation of the galactic disc, namely: $t_f \sim 2$ Gyr. This scenario enables to solve the problem of the lack of intergalactic gas infall. \end{abstract}
       
\begin {keywords}
Galaxy: fundamental parameters -- Galaxy: abundances -- galaxies: spiral -- galaxies: star formation.
\end{keywords}

       
\section{Introduction} 
It is recognized that the corotation resonance where the rotation velocities of galactic matter and density waves, responsible for spiral arms coincide, plays an important role in the galactic evolution and the value of the corotation radius is one of the fundamental galactic parameters. In literature one can find several methods of this quantity deriving in our Galaxy, for instance, by means of analysis of arms geometry, stellar or open clusters kinematics, HI and HII emissions (e.g., Lin et al. 1969; Marochnik et al. 1972; Mishurov \& Zenina 1999; L\'epine et al. 2001; Fernandez et al. 2001; Dias \& L\'epine 2005, etc.). 

In the present paper, we propose a new approach to evaluation of the corotation radius in our Galaxy. The method is based on the statistical analysis of oxygen radial distribution in the galactic disc (notice that Martin \& Roy 1995 and Scarano et al. 2010 also mentioned the importance of the corotation effects in analysis of abundance gradient in external galaxies). Oxygen was used since it is mainly produced by SNe II which are strongly concentrated in spiral arms. Hence it is the most pure indicator of spiral arms influence on the formation of radial abundance pattern in the galactic disc.

As an observed material, we use the data on oxygen distribution derived by Andrievsky et al. (2002 a,b,c) and Luck et al. (2003, 2006) over Cepheids. Being bright and very young objects with precise distances, these stars give reliable information about abundances of heavy elements, close to the one in interstellar medium, in the significant part of the galactic disc. 

The main finding of the above papers is that the radial distribution of metallicity in the galactic disc is {\it bimodal}, i.e. there is a rather steep gradient in the inner part of the disc at $5 \le r \le 7$ kpc and a plateau-like distribution for $r > 7$ kpc and up to about 10 kpc (the solar galactocentric distance $r_0 =7.9$ kpc). Hence, there is a bending in the slope of the distribution at $r \sim7$ kpc. 
       
The above fine structure in the radial abundance distribution is very important. It indicates that in the galactic disc the distribution formation is caused by some non-trivial process. Mishurov et al. (2002), Acharova et al. (2005; 2010) developed a theory of spiral arms influence on the radial distribution of oxygen in the galactic disc. They show that the bending in the slope of oxygen distribution is associated with the corotation resonance. However, in our previous papers we fitted the theory to the observations ``by eyes''. Now we propose a statistical method for the deriving of the corotation resonance location by means of analysis of oxygen distribution along the galactic radius. Simultaneously the so-called ``constant for the rate of oxygen synthesis'' is estimated.

       
\section {Statistical method}

Let $[X/H]^{ob}(r)$ be an observed distribution of any $X$ element along the galactic radius $r$ (as usual $[X/H] = log(N_X/N_H)_s - log(N_X/N_H)_{\odot}$, where $N_{X,H}$ is the number of $X$ element or hydrogen atoms in the object, the first item refers to a star located at the distance $r$, the second one - to the Sun). On the other hand, let us assume that $[X/H]^{th}(r)$ is a theoretical distribution of the corresponding element which depends on some vector $\hat{\lambda}$, the coordinates of the vector being the sought-for free parameters of our theory. To fit the observations we minimize the variance $\sigma^2$ over $\hat{\lambda}$:
       
\begin{equation}
\sigma^2(\hat\lambda) = \frac{1}{n-p}\sum_{i=1}^n([X/H]^{ob}_i - [X/H]^{th}_i)^2,
\end{equation}
here the summation is taken over all $i-th$ points of galactocentric radius where the abundances of the element were measured, $n$ is the number of observational data, $p$ is the number of the sought-for free parameters.
       
Unfortunately, $[X/H]^{th}(r)$ cannot be represented by any analytical or, at least, approximating formula. The theoretical abundance is derived as a numerical solution of a system of equations, the solution being dependent on a particular set of the free parameters. In our approach, there are 2 target free parameters ($p = 2$). To find them the following method is used: we solve numerically the equations of oxygen synthesis varying the coordinates of $\hat{\lambda}$ - vector within some regions and compute a net of $\sigma(\hat{\lambda})$. After that we construct the surface $\sigma$ as a function of the two free parameters and find the minimum ($\sigma_m = \min \sigma$) which determines the best set of the sought-for parameters.
       
To estimate the free parameters errors the confidence region was constructed: according to Draper \& Smith (1981) the confidence contour is determined by the intersection of the surface $\sigma(\hat{\lambda})$ with the horizontal plane at level:
       
\begin{equation}
\sigma^2_c = \sigma^2_m[1+\frac{p}{n-p}F(p,n-p,0.95)],
\end{equation}
where $F$ is Fisher's $F$-statistics and the 95\% level of confidence was adopted.

       
\section {Equations of oxygen synthesis}
       
The equations which describe the evolution of oxygen synthesis in the galactic disc are as follows (details see in Acharova et al. 2005 and 2010):
       
\begin{eqnarray}
\frac{\partial \mu_g}{\partial t}&=&-\frac{1}{r}\frac{\partial}{\partial r}(r u\mu_g)+ f-\psi +\nonumber\\
&&\int\limits_{m_L}^{m_U}{(m-m_w)\psi(t-\tau_m)\phi (m)\,dm},
\end{eqnarray}

\begin{equation}
\frac{\partial \mu_s}{\partial t}=\psi - \int\limits_{m_L}^{m_U}{m\psi(t-\tau_m)\phi (m)\,dm},
\end{equation} 

\begin{eqnarray}
\frac{\partial \mu_O}{\partial t}&=&\int\limits_{m_L}^{m_U}{(m-m_w)\,Z(t-\tau_m)\psi(t-\tau_m)\phi (m)\,dm} + \nonumber\\
&&E + fZ_{f}-Z\psi- 
\frac {1}{r}\frac {\partial {}}{\partial r}\left(ru\mu_O - r\mu_g D\frac {\partial Z}{\partial r}\right),
\end{eqnarray}
where
$\mu_{g,s}$  are surface densities correspondingly for interstellar gas and stars,
$\mu_O$ is the density of oxygen in interstellar medium,
$Z=\mu_O/\mu_g$ is the fraction of oxygen (hence $[O/H] = log (Z/Z_{\odot}$)), 
$\psi$ is the star formation rate (SFR; $\psi=\nu\mu_g^{1.5}$, $\nu$ is a normalizing coefficient),
$\phi(m)$ is Salpeter's initial mass function with the exponent of -~2.35 (stellar masses $m$ are in solar units),
$E$ is the rate of oxygen synthesis,
$f$ is the infall rate of intergalactic gas on to the galactic disc,
$u$ is the radial velocity of gas within the galactic disc averaged over the galactic azimuth,
$t$ is time (in Gyr),
$\tau_m$ is the life-time of a star of mass $m$ on the main sequence,
$m_L=0.1$, $m_U=70$,
$m_w$ is the mass of stellar remnants (white dwarfs, neutron stars, black holes).

The infall rate of intergalactic gas on to the galactic disc is described as $f=A\exp(-r/r_d-t/t_f)$ with the radial scale 
$r_d = 3.5$ kpc (Marcon-Uchida et al. 2010) which is an intermediate value between $r_d = 2.5$ and $4.5$ kpc usually used in galactic nuclear synthesis modeling (e.g., Naab \& Ostriker 2006; Fu et al. 2009; Sch\"{o}nrich \& Binney2009). Constants $t_f$ and $A$ are defined below.

We made experiments with various abundances of the infall gas from $Z_f=0.02\,Z_{\odot}$ to $0.1\,Z_{\odot}$. 
In accordance with Lacey \& Fall (1985) our final abundances weakly depend on the exact value of $Z_f$ if 
$Z_f \le 0.1\,Z_{\odot}$. Below we demonstrate the results for $Z_f=0.02\,Z_{\odot}$ 
which is slightly less than the mean content of heavy elements in halo stars ($\sim 0.03\,Z_{\odot}$, Prantzos 2008).

The enrichment rate of interstellar medium by oxygen is represented as follows:
$E=\eta P R$, where $P=2.47$ is the mass (in solar units) of ejected oxygen per one SN\,II explosion 
(Tsujimoto et al. 1995), $R$ is the rate of SNe\,II events 
$$R(r,t)=0.9975\int\limits_{8}^{m_U}{\psi(r,\,t-\tau_m)\phi (m)\,dm},$$ 
the factor $\eta$ describes the influence of spiral arms on radial oxygen pattern formation and is represented by 
the expression: $\eta=\beta |\Omega(r)-\Omega_P|\Theta$, where $\Omega(r)$ is the angular rotation velocity of galactic disc,
$\Omega_P$ is the angular rotation velocity of spiral density waves (recall that whereas
the galactic matter rotates differentially, i.e. $\Omega(r)$ is a function of $r$, galactic density waves rotate
as a rigid body, $\Omega_P = const$, the corotation radius $r_C$ is determined from equation $\Omega(r_C) = \Omega_P$),
$\beta$  is the constant of the rate of oxygen 
synthesis, $\Theta$ is a cut-off factor: $\Theta(r) = 1$ if $r$ is contained between the inner and outer Lindblad resonances and $\Theta = 0$ otherwise. In our computations, we use the rotation curve based on Clemens (1985) data 
adjusted for the scale $r_0 = 7.9$ kpc.
       
The last term in parentheses in Equation (5) describes the turbulent radial diffusion of heavy elements. To derive the diffusion coefficient $D$ we use the simple gas kinetic approximation (Mishurov et al. 2002).

The above system of equations may be divided into 2 groups: {\it i)} equations (3,4) which describe galactic disc formation, and {\it ii)} the ``chemical'' equation (5) of oxygen synthesis. The target 2 free parameters, $\Omega_P$ and $\beta$, enter only the equation (5). Hence, the first group of equations can be solved independently of the last equation. Therefore, the algorithm of solving the above equations is as follows. 
       
At first step, independently of Equation (5) we solve equations (3,4) by means of Lelevier method (Potter 1973) and find the time evolution of the radial profiles of gaseous and stellar densities, $\mu_g(r,t)$ and $\mu_s(r,t)$ (simultaneously SFR, $\psi(r,t)$ is derived). To perform this part of modelling, we have to specify a value of the time-scale of gas infall on to the galactic disc, $t_f$, and adopt an expression for $u(r)$. After that, the constants $\nu$ and $A$, entering these group of equations, are obtained by fitting the normalising conditions that at present epoch 
($t = T_D = 10$ Gyr - the age of galactic disc) and solar position stellar and gaseous densities have to be equal to:
$\mu_s(r_0,T_D) = 40$ M$_{\odot}\,pc^{-2}$ and 
$\mu_g(r_0,T_D) = 10$ M$_{\odot}\,pc^{-2}$ (Haywood et al. 1997). Taking into account that the part of stellar remnants is of the order of 10\% of $\mu_s$ 
(our experiments confirm this supposition) we have that at present time and solar galactocentric distance summary density has to be about 54 M$_{\odot}$ pc$^{-2}$ - a value close to Marcon-Uchida et al. 2010. However, below we also discuss other normalising conditions.

At the second step, we solve the chemical equation (5) for a set of $\Omega_P$ and $\beta$ (as boundary conditions we demand that the solution has to be finite at the galactic centre and the galactic disc edge which we dispose at $r = R_G = 35$ kpc). Repeated this procedure we find the minimum of $\sigma = \sigma_m$, hence estimate the best values for $\Omega_P$ and $\beta$, then compute their errors as it was described in Sec. 2. At last we derive the location of the corotation resonance $r_C$.

\section {Step 1: galactic disc formation}

In what follows we examine 2 limiting cases for $t_f$: $t_f = 2$ Gyrs -- {\it rapid disc formation}, and 
$t_f = 7$ Gyrs -- {\it slow disc formation}. For the radial gas flow 3 models were studied.
       \footnote {Strictly speaking in a full formulation of the problem we have to solve simultaneously gas dynamical equations which take into account the effects of the central bar, spiral arms, etc. However, such complicated task is still beyond the scope of approach usually used in theories of nuclear synthesis in galactic discs. Instead of that, following Lacey \& Fall (1985), Portinari \& Chiosi (2000) we examine several model representations for the radial velocity $u$.}
       
1) No radial flow of gas: $u = 0$ ({\it models M20 and M70} correspondingly for the above two rates of disc formation). 
       
2) Gas radial inflow: 
{\it case a)}~$u = -0.3r$ km s$^{-1}$, if $r < 3.3$ kpc and $u = -1$ km s$^{-1}$, if $r \ge 3.3$ kpc ({\it models M2a and M7a}); 
{\it case b)}~$u = -0.1r$ km s$^{-1}$ ({\it models M2b and M7b}); 
       
3) Gas radial outflow: $u = +0.5$ km s$^{-1}$ ({\it models M2O and M7O}). 
       
In Fig.1 evolution of gaseous radial profiles with time is shown. In significant part of galactic radii current gaseous density distributions happen to be close to the observed ones and profiles derived by Portinari \& Chiosi (2000),
Fu et al. (2009), Naab \& Ostriker (2006) and Sch\"{o}nrich \& Binney (2009) . Comparing the distributions we cannot prefer a certain model only on the basis of density profile (models M2O and M7O have holes in the galactic centre but they do not describe oxygen radial distribution well).

\begin{figure}
\includegraphics {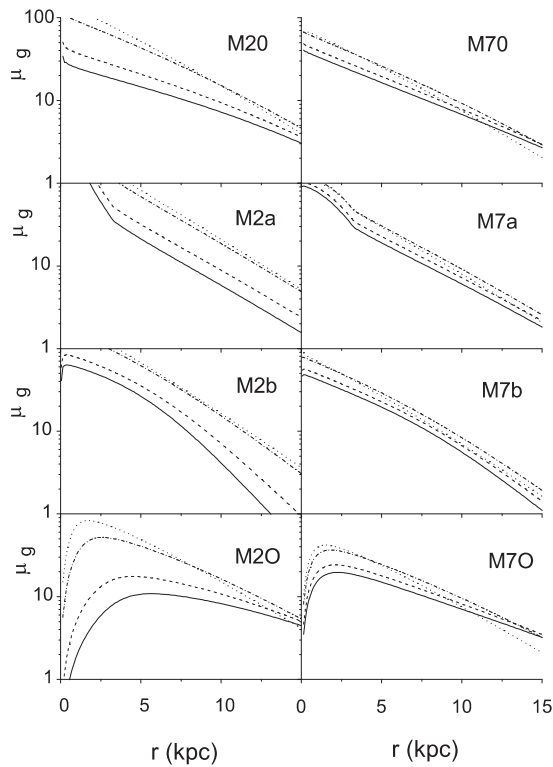}
\caption{Evolution with time of the profiles of gaseous density: {\it dotted line} - for $t=2$; 
{\it dashed - dotted line} - for $t = 4$; {\it dashed line} - for $t=8$; {\it solid line} for 
$t=10$ Gyr.}
\label{f1}
\end{figure}
       
So, we need some additional arguments which enable us to make a choice among various scenarios of galactic disc formation. Such information comes from observations: according to Sancisi \& Fraternali (2008) and Bregman 
(2009) current rate of total (integrated over the galactic disc) gaseous mass falling on to the galactic disc is of the order of 0.1 - 0.2 M$_{\odot}$ yr$^{-1}$ whereas the total star formation rate is $\sim 1-2$ M$_{\odot}$ yr$^{-1}$ 
(Robitaille \& Whitney 2010) although one can meet a higher level of SFR $\sim 5$ M$_{\odot}$ yr$^{-1}$ (Hartmann 2007). The above values may be considered as global normalisations which are complementary to the particular ones at solar position.
       
In Table 1 we give the computed total rate of forming stars, 
$\dot M_s=2\pi\int_0^{R_G}\psi rdr$, and infall gas mass, 
$\dot M_f=2\pi\int_0^{R_G}frdr$ at the present epoch. From the Table one can see that only models of rapid galactic disc formation ($t_f \sim 2$ Gyr) demonstrate low current infall rate, $\dot M_f \sim 0.1 -0.2$ M$_{\odot}$ yr$^{-1}$, and 
SFR at level $\dot M_s \sim 1.5$ M$_{\odot}$ yr$^{-1}$, the both quantities fit the above observed values. Scenarios of slow disc formation ($t_f \sim 7$ Gyr) lead to greater current SFR, $\dot M_s \sim 3 - 4$ M$_{\odot}$ yr$^{-1}$, but they require the infall rate of the order of magnitude higher than that observed by Sancisi \& Fraternali (2008) and Bregman (2009). Obviously, one should be prudent in such estimates since future observations could discover such a 
huge amount of baryonic matter falling on to the galactic disc. But at the moment we do not have such data, at least for our 
Galaxy.

       
\begin{table}
\caption[]{Constants $A$ (M$_{\odot}$ pc$^{-2}$ Gyr$^{-1}$) and $\nu$ (M$_{\odot}^{-0.5}$ pc Gyr$^{-1}$); integrated rates of gas infall, $\dot M_f$, and star formation rate, $\dot M_s$ (M$_{\odot}$ yr$^{-1}$) and  minimal $\sigma_m$ derived in various models\\}
\begin{tabular}{|l|c|c|c|c|c|c|c|c}       
\hline
& M20  & M70&  M2a & M7a & M2b & M7b & M2O & M7O \\
\hline
$A$          & 255.0&97.0& 384.8&122.1&303.0& 98.8&230.9&89.8 \\
$\nu$        & 0.070&0.14& 0.07&0.13 &0.058& 0.14&0.068&0.14 \\
\hline
$\dot M_f$   & 0.13 & 1.74 & 0.19 & 2.19 & 0.15 & 1.77 & 0.12 & 1.61 \\
$\dot M_s$   & 1.40 & 3.00 & 1.75 & 3.70 & 1.75 & 3.06 & 1.33 & 2.77 \\
\hline
$\sigma_m$   &0.050&0.072&0.060&0.066&0.050&0.043&0.086&0.068\\
\hline
\end{tabular}
\end{table}

Let us look at the problem from another angle. There is a belief that star formation is mainly fuelled by the gas infall on to the galactic disc. Hence, the total rates of gas infall and star formation should be close to each other in order to the interstellar gas not to be consumed due to star formation. But as we wrote above, at present epoch in our Galaxy the total star formation rate is at least 5 times higher than the one of gas infall. 
       
To explain this problem, in Fig. 2 we show evolution of the ratio of gas infall rate on to the galactic disc to star formation rate, $f/\psi$, with time as a function of $r$ for 4 models. From the Figure it is seen that, at each moment of time there is a region in the galactic disc where $f/\psi > 1$. This means that the interstellar gas is not fully consumed in the Galaxy (from Fig. 1 one can also see that the gaseous density is not equal to zero at any galactocentric radius).

Besides, as it was expected, in scenarios with gas radial inflow the region, where $f/\psi > 1$, is shifted closer to the galactic centre. Indeed, gaseous density decreases with galactocentric radius, star formation rate also decreases with $r$. Hence, due to interstellar gas radial inflow the region with depressed star formation rate moves closer to the galactic centre and this results in the ratio $f/\psi$ increase. In addition gas radial inflow replenishes the inner part of the galactic disc with building material for star formation fuelling. 
So, as we can see, at each 
epoch of galactic evolution the reserve of gas forms into the galactic disc and it fuels star formation process. 
\begin{figure}
\includegraphics {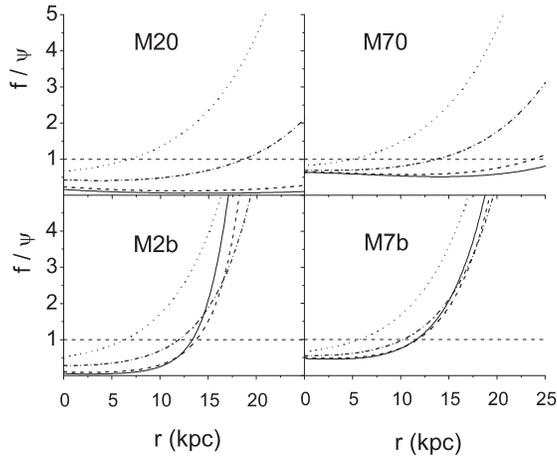}
\caption{Evolution with time of the ratio of infall rate to SFR (line types correspond to the time moments as in Fig. 1). Horizontal dashed lines separate regions in the Galaxy where $f > \psi$. Here a reserve of gas is formed for the future star formation.}
\label{f2}
\end{figure}

How should we change the normalisations of the current gaseous and stellar densities at solar distance in order to obtain the global star formation rate at level, say, $\dot M_s \sim 5$ M$_{\odot}$ yr$^{-1}$? Our experiments give the following results. In scenario of rapid galactic disc formation ($t_f\sim 2$ Gyr) and no gas radial inflow, gaseous and stellar densities happen to be 
unrealistically large: $\mu_s(r_0,T_D) \sim 155$ M$_{\odot}$ pc$^{-2}$; $\mu_g(r_0,T_D) \sim 30$ M$_{\odot}$ pc$^{-2}$ albeit the rate of gas infall $\dot M_f \sim 0.5$ M$_{\odot}$ yr$^{-1}$ (in case of gas radial inflow like in models ``b'' $\mu_s(r_0,T_D)\sim$60 M$_{\odot}$ pc$^{-2}$; $\dot M_f \sim 0.3$ M$_{\odot}$ yr$^{-1}$ however $\mu_g(r_0,T_D)$ keeps at very high level $\sim 30$ M$_{\odot}$ pc$^{-2}$). In scenario of slow galactic disc formation ($t_f\sim 7$ Gyr) and no gas radial inflow we have: $\mu_s(r_0,T_D) = 75$ M$_{\odot}$ pc$^{-2}$; $\mu_g(r_0,T_D) = 14$ M$_{\odot}$ pc$^{-2}$ but the rate of gas infall $\dot M_f \sim 3.1$ M$_{\odot}$ yr$^{-1}$ is of the order of magnitude higher than the observed value (if we take into account gas radial inflow like in models ``b'' we derive: $\mu_g(r_0,T_D) \sim 25$ M$_{\odot}$ pc$^{-2}$; $\mu_s(r_0,T_D) \sim 40$ M$_{\odot}$ pc$^{-2}$ but 
$\dot M_f \sim 2.6$ M$_{\odot}$ yr$^{-1}$ again is too high). Perhaps such situation can be implemented in some galaxies but it is not supported by observations for our Galaxy.

There is one additional argument in favour of rapid disc formation scenario: according to Allende Prieto (2010) and Zasov \& Sil'chenko (2010) galactic discs were formed rapidly, at least last 8 Gyr their sizes did not change significantly.

\section {Step 2: estimate $\Omega_P$, $r_C$ and $\beta$}
       
Despite the conclusion made in the previous Section below we give the results for all the above models. So, at the next step we solve the chemical Equation (5), compute the variance $\sigma^2$ and find its minimum which determines the best values of $\Omega_P$ and $\beta$ (to catch the minimum of the variance we varied $\Omega_P$ and $\beta$ in a wide region: $10 \le \Omega_P \le 60$ km s$^{-1}$ kpc$^{-1}$; $0 \le \beta \le 0.6$ Gyr.).

As an observed material we use the data of oxygen distribution along the galactic radius, derived over Cepheids, which were averaged within bins of 0.5 kpc width (see Acharova et al. 2005; 2010). The corresponding distribution is drawn in Fig. 3 at step 0.25 kpc in the region from 5.25 up to 10 kpc (in total, the number of observed points is $n=20$). The bars in the Figure are the rms scatters of the mean abundance within the bin. 
\begin{figure}
\includegraphics {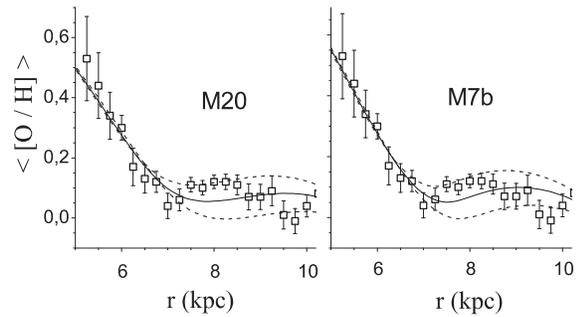}
\caption{ Comparison of theoretical and observed oxygen radial distributions for models M20 and M7b. {\it Solid lines} are for the best values of $\Omega_P$ and $\beta$, {\it dashed lines} -- for ($\Omega_P + \Delta_{\Omega_P}; \beta + \Delta_{\beta}$) and ($\Omega_P - \Delta_{\Omega_P}; \beta - \Delta_{\beta}$).} 
\label{f3}
\end{figure}
To illustrate our method, in Fig. 4, we show an example of the surface $\sigma (\Omega_P, \beta)$ for model M20. 
\begin{figure}
\includegraphics {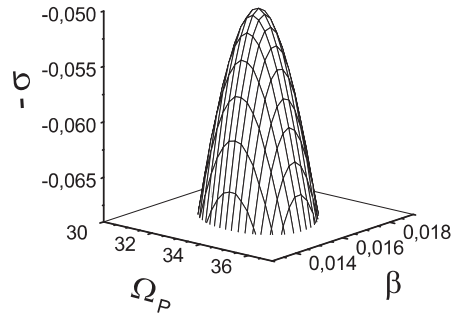}
\caption{An example of the surface $-\sigma (\Omega_P , \beta )$ in case of M20. For better visualisation we draw the figure ``bottom up''. The best estimates for the parameters are: $\Omega_P = 33.2$ km s$^{-1}$ kpc$^{-1}$; 
$\beta = 0.016$ Gyr.} 
\label{f4}
\end{figure}
To estimate the errors of the sought-for free parameters, we compute the section of the above surface by the horizontal plane at $\sigma \equiv \sigma_c = 1.181 \sigma_m$ (see equation (2), in our case the value of Fisher's statistics happens to be $F(2,18,0.95) \approx 3.55$, Draper \& Smith 1981). The section is demonstrated in Fig. 5 by the shaded figure for model M20. Its ellipse-like border delineates the confidence region. The semi-axes of the ellipse give the errors. The results of variance computations are shown in Table 1. Consider them in more detail. 
\begin{figure}
\includegraphics {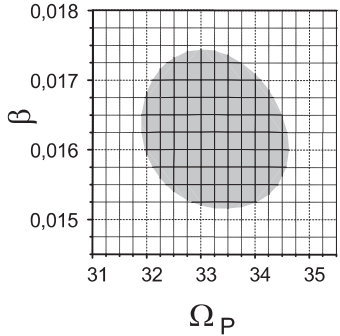}
\caption{The shaded region is an intersection of the surface $\sigma(\Omega_P, \beta)$ from Fig. 4 by the horizontal plane $\sigma = \sigma_c = 0.059$. This region determines the confidence one at probability level 95 \% (see Sec. 2). 
Its border gives the estimate of the parameters errors: $\Delta_{\beta} = \pm 0.001$; $\Delta_{\Omega_P} = \pm 1.4$ km s$^{-1}$ kpc$^{-1}$.} 
\label{f5}
\end{figure}

First of all, by means of Fisher's statistics analysis the significance of gas radial inflow inclusion into consideration can be estimated. To do that we have to compare models ``a'',``b'' and ``O'' with the ones without gas radial inflow separately for the above two scenarios of galactic disc formation. The result is as follows. 

{\it i}) In scenario of slow disc formation the best model is M7b. Other models have significantly higher values of $\sigma_m$ and they are to be rejected.

{\it ii}) In scenario of rapid disc formation two models, M20 and M2b, have the same values of $\sigma_m = 0.050$, other ones should be rejected. However, we cannot decide between M20 and M2b: the null hypothesis ($u = 0$) cannot be rejected. In other words, in this scenario inclusion of the gas inflow is insignificant.

So, below we discuss two models - M20 and M7b. Let us compare them and test the null hypothesis that the corresponding variances do not differ significantly against the alternative one that they do. For this, according to Martin (1971) we should check if the value $F=(min \, \sigma_{M20}/min \, \sigma_{M7b})^2$ lies in the range: $1/F_{\alpha/2} < F < F_{\alpha/2}$ ($\alpha$ is the significance level and we adopt that the degrees of freedom are approximately the same in the both alternatives). From Table 1 we derive $F \approx 1.35$, whereas for $\alpha = 0.02$ $F_{\alpha/2} = F(18,18,0.99) \approx 3.14$ (Draper \& Smith 1981). Hence, the above inequality is satisfied and we conclude: the difference between models M20 and M7b is undetected statistically (on the basis of available observational data of oxygen radial distribution and a rather simple theoretical approach). At the moment, the only way to discriminate between them is as it was discussed in Sec. 4. Finally, our estimates for the target parameters are as follows:
       
1) {\it M20}: $\Omega_P = 33.2 \pm 1.4$ km s$^{-1}$ kpc$^{-1}$ (the corotation radius $r_C = 7.0 \pm 0.3$ km s$^{-1}$ kpc$^{-1}$); $\beta = 0.016 \pm 0.001$ Gyr.
       
2) {\it M7b}: $\Omega_P = 30.5 \pm 1$ km s$^{-1}$ kpc$^{-1}$ (the corotation radius $r_C = 7.6^{+0.3}_{-0.2}$ km s$^{-1}$ kpc$^{-1}$); $\beta = 0.036 \pm 0.002$ Gyr.
       
In Fig. 3 the best theoretical oxygen distributions for the above two models and 
the distributions for the upper and lower values of $\Omega_P$ and $\beta$ are shown. From the figure it is seen that our theory explains 
the bimodal radial distribution of oxygen in the disc of our Galaxy rather well.

\section {Discussion}
       
In the present paper, a new method for evaluation of the corotation radius in the Galaxy is developed. Our approach is based on a statistical analysis of the bimodal structure of oxygen radial distribution in the galactic disc determined over Cepheids. By means of treatment of observational data we derive that the corotation resonance happens to be situated at $r_C \sim 7.0 - 7.6$ kpc depending on the rate of intergalactic gas infall on to the galactic disc (the statistical error is $\sim$0.3 kpc). The above value for the corotation radius is close to the solar galactocentric distance $r_0 = 7.9$ kpc. Simultaneously the constant for the rate of oxygen synthesis in the galactic disc $\beta \sim 0.016 - 0.036$ Gyrs was obtained. 
       
We also argue in favour of a short time-scale formation of the galactic disc, namely: $t_f \sim 2$ Gyrs. This scenario enables to solve the problem of the lack of intergalactic gas infall, i.e. the very low present integrated rate of gas infall on to the disc, $\dot M_f \sim 0.1 - 0.2$ M$_{\odot}$ year$^{-1}$ observed by Sancisi \& Fraternali (2008) and Bregman (2009), whereas the integrated star formation rate is expected to be $\dot M_s \sim 1 - 2$ M$_{\odot}$ year$^{-1}$ (Robitaille \& Whitney 2010). Higher level of the current global star formation rate, $\dot M_s \sim 5$ M$_{\odot}$ year$^{-1}$, needs too high rate of gas infall on to the galactic disc or unrealistically high gaseous and stellar current densities at the solar galactocentric distance.
      
\section*{Acknowledgments}
       
Authors thank to referee for vary important comments, to Profs. Yu.Shchekinov and A.Zasov for helpful discussions. The work was supported in part by grants No. 02.740.0247 and P685 of Federal agency for science and innovations. IAA thanks to the Russian funds for basic research, grant No. 11-02-90702.
      

\label{lastpage}
       
\end{document}